\documentclass{blois}

\def\Journal#1#2#3#4{{#1} {\bf #2}, #3 (#4)}

\def\PLB{{\em Phys. Lett.}  B}
\def\PRL{\em Phys. Rev. Lett.}
\def\PRD{{\em Phys. Rev.} D}

\def\be{\begin{equation}}
\def\ee{\end{equation}}
\def\bea{\begin{eqnarray}}
\def\eea{\end{eqnarray}}

\newcommand{\Photo}{}

\begin{document}
\vspace*{4cm}

\title{Time-space symmetry as a solution to the mass hierarchy
\\of charged lepton generations
}

\author{ Vo Van Thuan }

\address{Vietnam Atomic Energy Institute (VINATOM)\\
59 Ly Thuong Kiet street, Hoan Kiem district, Hanoi, Vietnam
\\Email: vvthuan@vinatom.gov.vn}

\maketitle\abstracts{
Based on an extended time-space symmetry, a cylindrical model of gravitational geometrical dynamics with two time-like extra-dimensions leads to a microscopic geodesic description of the curved space-time. Due to interaction of a Higgs-like cosmological potential with individual space-time fluctuations, the original time-space symmetry is spontaneously broken, inducing a strong time-like curvature and a weak space-like deviation curve. As a result, the basic Klein-Gordon-Fock equation of a free massive elementary particle was derived, which implies a duality between the quantum mechanics equation and a microscopic geodesic description in the frame of general relativity. Consequently, Heisenberg inequalities are determined explicitly by the space-time curvatures. Moreover, extending curvatures to higher time-like dimensional hyper-spherical surfaces than one of the basic common cylindrical configuration, we found reasonable mass ratios of all charged leptons and succeeded to fix the number of their generations to be three. Following to concepts of the standard cosmological model, a possible experimental verification of mass ratio variation is proposed.}

\section{Introduction}

 Kaluza and Klein ~\cite{Ka1}$^{,}$~\cite{Kl1} were pioneers to propose a space-like extradimension (ED) which is to compact in a micro circle in a relation to general relativity. Klein and Fock ~\cite{Kl1}$^{,}$~\cite{Fo1} also found a formalism that the equation of motion of a massive particle in 4D space-time can be obtained by reducing the EDs of a massless particle in a higher dimensional time-space. For the semi-classical approach to quantum mechanics introduced by de Broglie and Bohm ~\cite{Br1}$^{,}$~\cite{Bo1} the hidden parameters are somehow reminiscent of EDs. Later on, the evidence for violation of Bell inequalities ~\cite{Be1}$^{,}$~\cite{Fr1} abandoned the models with local hidden parameters, however, leaving the door open to non-local hidden variables. In the wake of high dimensional superstring models, another trend has been developed following the Kaluza-Klein geometrical dynamics, of which most applied space-like EDs, while few others considered time-like ones. There are two main approaches with time-like EDs: membrane models in the Anti-de-Sitter geometry (AdS), such as ~\cite{Ma1}$^{,}$~\cite{Ra1}and induced matter models~\cite{We1}$^{,}$~\cite{We2}. In particular, Maldacena ~\cite{Ma1} found a duality between AdS and conformal fields as AdS/CFT formalism. Randall and Sundrum ~\cite{Ra1} applied an infinite 5D AdS model for a hierarchy solution. For the induced matter approach, Wesson~\cite{We1} has proposed a space-time-matter model describing proper mass as a time-like ED. A geometrical dynamic model for elementary particles was proposed by Koch~\cite{Ko1}$^{,}$~\cite{Ko2} with a time-like ED which offered a method for derivation of Klein-Gordon equation. Our preliminary study ~\cite{Vo1} following the induced-matter approach was based on the time-space symmetry in which the Klein-Fock reduction formalism was used and the two time-like extra-dimensions are made explicit in terms of the quantum wave function $\psi$ and the proper time variable $t_0$. For a next step in ~\cite{Vo2} it was proved in a more direct way that the quantum wave equations in 4D space-time can be identical to a general relativistic geodesic description of curved extradimensional time-space and, as a result,  the Heisenberg indeterminism is shown to originate from the space-time curvatures. In the present study we do a new attempt to apply the extended time-like curvature to solve the problem of mass hierarchy of charge lepton generations and to prove a significance of time-space symmetry.

\section{Time-space symmetry with time-like extradimensions}

In ~\cite{Vo1}$^{,}$~\cite{Vo2}, we constructed an ideal 6D flat extended symmetrical time-space $\{t_1,t_2,t_3 \mid x_1,x_2,x_3\}$:
 \begin{equation}
dS^2=d\vec{k}^2-d\vec{l}^2=dt_k^2-dx_l^2;
\label{eq1}
\end{equation}
 and carried out the investigation on the time-space symmetrical "lightcone" embedded in the 6D flat time-space $(\ref{eq1})$:

\begin{equation}
d\vec{k}^2=d\vec{l}^2\qquad \rightarrow\qquad dt_k^2=dx_l^2;
\label{eq2}
\end{equation}

Where $k,l=1 \div 3$ are summation indexes that: $d\vec{k}^2={\sum dt_k}^2$ and $d\vec{l}^2={\sum dx_l}^2$. Hereafter, natural units ($\hbar=c=1$) are used generally unless it needs an explicit quantum dimension. Any differentials-displacements of $dt_k$  and $dx_l$ correspondingly in 3D-time and in 3D-space can be independent from each other. They are naturally considered as the most primitive sources of physical potential energy in the symmetrical time-space. The time-like source can form a special global vacuum potential in 3D-time equivalent to the original tachyonic Higgs potential. Another half of the primitive source in 3D-space would contribute to the global cosmological constant $\Lambda$. We assume a postulate that the equality $(\ref{eq2})$  between the squared linear time-like and space-like intervals $d\vec{k}^2=d\vec{l}^2$ is to accept as a fundamental time-space symmetry not only for Euclidean geometry, but also to be conserved for the linear translational elements of more generalized space-time geometries and in their transformation from a higher dimensional geometry to lower one. This assumption denoted as a principal conservation of linear translation (CLT) bases on our experience of the firm Lorentz invariance as well as of the homogeneity and isotropy of 4D Minkowski space-time.
\\Let us select among differentials $(\ref{eq2})$ those displacements for which a time-like displacement correlates with a space-like one by a harmonic function $f(t_k,x_l)$.  In the flat 6D time-space we introduce the following 6D isotropic plane wave equation:

\begin{equation}
\frac{\partial^2f }{\partial t_k^2}=\frac{\partial^2f }{\partial x_l^2};
\label{eq3}
\end{equation}

Where $t_k$  and $x_l$ remain Descartes coordinates, describing transmission of plane waves $f(t_k,x_l)$ in Euclidean time-space. The 6D wave transmission $(\ref{eq3})$ can serve as a primitive energy-momentum formation of physical objects, symmetrical in 3D-time and 3D-space. Most of the primary form of energy in 3D-time probably is almost unobservable from our 3D-space and would be a kind of dark energy. Zeldovich in~\cite{Ze1} applied a special global cosmological vacuum to a  $\Lambda$-model for elementary particles.  Qualitatively, following~\cite{Ze1}, we would propose a scenario of formation of a single direction of time evolution of microscopic substances, using the time-like vacuum potential of an effective time-like "cosmological constant" $\Lambda_T$. Namely, the global potential in 3D-time is able to generate strong quantum fluctuations in space-time. In particular, there are individual fluctuations being able to fix a time-like circular polarization in 3D-time equivalent to breaking of space-time symmetry and similar to the Higgs mechanism to induce the proper mass to a kind of identical elementary particles. The plane wave $(\ref{eq3})$ acquires a time-like circular polarization namely along the $t_3$ axis, keeping strictly an arrow evolution from the past to the future and being constrained by a time-like cylindrical condition. Polar coordinates $\{\psi(t_0),\varphi(t_0),t_3\}$ are used for the 3D-time instead of linear coordinates $\{t_k\}$ where vector $dt_0$ is evolving along a circle located in a plane orthogonal to the longitudinal vector $dt_3$. The squared differential along the time curve reads:

\begin{equation}
dt^2=d\psi(t_0)^2+\psi^2d\varphi(t_0)^2+dt_3^2=ds^2+dt_3^2;
\label{eq4}
\end{equation}

Where $ds$ is the Lorentz invariant interval characterizing the curvature term of time $t$, while the only linear term of time $dt_3$ in $(\ref{eq4})$ is getting identical to the time-like interval $d\vec{k}$ in  $(\ref{eq2})$, in according to the CLT principle. The direction of vector $dt_0$ serves a label attached to a particle or an anti-particle. Vectors $dt_3$ and $dt_0$ form a local basis for their vector summation $dt$ such as $\Omega dt=\Omega_0dt_0 + \Omega_3dt_3$, where $\Omega$, $\Omega_0$ and $\Omega_3$ are linear scale factorization parameters. Therefore, the local time-like geodesic $dt$ in 3D-time plays a role of a real time duration in our 3D-space.
\\Linear translation of a material point in 3D-space is taken along a linear vector $d\vec{l}$ defined along the same direction of the linear vector $d\vec{l}$ in $(\ref{eq2})$. Simultaneously, the material point can also rotate around an arbitrary 3D-spatial direction characterized by a spin $\vec{s}$. Originally, vector $\vec{s}$ in 3D-space is described by a spherical system attached to the material point $\{\psi(x_n),\theta(x_n),\varphi(x_n)\}$, where $\psi$ is the deviation variable extended in 3D-space, $\varphi$ is the azimuth around the spherical axis $\vec{l}$, while $\theta$ as the zenith between $\vec{l}$ and $\vec{\psi}$ defines its right-additive as the angle between $\vec{s}$ and $\vec{l}$. For conserving Lorentz invariance, we are restricted to consider the projection $s_n$ of $\vec{s}$ on axis $x_l$ which defines a local rotation in a plane $P_n$ orthogonal to vector $dx_l$ and the local proper coordinate $x_n\subset P_n$ now plays a role of a new affine parameter in 3D-space. The squared differential along the spatial curve reads:

\begin{equation}
d\lambda^2=d\psi(x_n)^2+\psi(x_n)^2\left [ d\theta^2+sin^2\theta d\varphi(x_n)^2\right]+dx_l^2=d\sigma^2+dl^2;
\label{eq5}
\end{equation}

Where $d{\sigma}^2=d\sigma_{ev}^2+d\sigma_{od}^2$ is a formal local interval in 3D-space; $d\sigma_{ev}$ is the P-even component related to rotation with the spatial symmetry and $d\sigma_{od}$ is the P-odd component implying a contribution of parity nonconservative (PNC) rotation. In general, the P-even term may contain a non-Lorentz invariant component of longitudinal fluctuations, however, the latter being compensated itself due to its forward-backward symmetry that does not contribute explicitly to our consideration.
\\In the extended time-space, $\psi=\psi(t_0,t_3,x_n,x_l)$ and $\varphi=\Omega_0t_0+\Omega_3t_3-k_n x_n-k_l x_l=\Omega t-k_jx_j;$ where $\{x_j\}$ are general spatial coordinates, describing both linear translation and rotation. Physical meaning of $\Omega$, $\Omega_0$ and $\Omega_3$ is defined as of time-like rotational velocities, while $k_j$, $k_n$ and $k_l$ are spatial wave factors. This procedure implies that two independent EDs $\psi$ and $\varphi$ turn into the dynamical parameters depending on other 4D space-time dimensions. Again, based on the CLT principle, the 6D "lightcone" geometry $(\ref{eq2})$ being generalized with curvature is turning into a new quadratic form:

\begin{equation}
dt^2-ds^2={dt_3}^2=dl^2=d\lambda^2-{d\sigma}^2;
\label{eq6}
\end{equation}

Accordingly, the 6D time-space representation is transformed to a 4D space-time geometry of a spinning particle describing both translation and rotation:

\begin{equation}
{d\Sigma}^2=ds^2-{d\sigma}^2=dt^2-d\lambda^2;
\label{eq7}
\end{equation}

Where $d\Sigma$ is a total 4D interval which is to be fixed as a Lorentz  invariant only after a transformation from the 6D time-space to the 4D space-time. The more generalized quadratic form $(\ref{eq7})$ includes, as its partial linearization case, the traditional quadratic form of 4D Minkowski space-time describing the linear translation:
  \begin{equation}
ds^2-d\sigma_{od}^2\approx ds^2=dt^2-dl^2;
\label{eq8}
\end{equation}

Where $d\sigma_{od}$ is too small as originated from PNC effect and usually may be ignored. Equation $(\ref{eq8})$ proves the consistency of geometry $(\ref{eq7})$ with special relativity.

\section{Geodesic description or microscopic gravitational waves}
\label{sec:Geodesic}

 In according to transformation from 6D time-space to 4D space-time we derived in ~\cite{Vo2} a geodesic acceleration equation of deviation $\psi$. The 3D local affine parameters in this equation are the transverse evolving time $t_0$ (proper time) and the transverse rotational space-like variables $x_n$ (proper spinning). In according to time-space symmetry $(\ref{eq2})$ we postulate that any deviation from the linear translation in 3D-time is to compensate by a deviation in 3D-space, i.e. a balance established between two extended differentials: $Du(t_0)=Du(x_n);$ where the velocity $u(s)=\frac{\partial \psi}{\partial s}$. Therefore, the symmetrical equation of geodesic acceleration of deviation $\psi$ in both 3D-time and 3D-space reads:

\begin{equation}
\frac{\partial^2 \psi}{\partial {t_0}^2}+\Gamma _{\alpha \beta }^{\psi }\left ( \frac{\partial t_\alpha }{\partial t_0} \right )\left ( \frac{\partial t_\beta }{\partial t_0} \right )=\frac{\partial^2 \psi}{\partial {x_n}^2}+\Gamma _{\gamma \sigma}^{\psi }\left ( \frac{\partial x_\gamma }{\partial x_n} \right )\left ( \frac{\partial x_\sigma }{\partial x_n} \right );
\label{eq9}
\end{equation}

Where $t_\alpha,t_\beta\in \{\psi(t_0),\varphi (t_0),t_3\}$ and $x_\gamma, x_\sigma \in \{\psi(x_n), \varphi(x_n),x_l \}; l,n=1\div3$. In both sides of $(\ref{eq9})$ among the related Christoffel symbols except  $\Gamma_{\varphi(t_0)\varphi(t_0)}^\psi=-\psi$  and $\Gamma_{\varphi(x_n) \varphi(x_n)}^\psi=-\psi.sin^2⁡\theta$  , all other terms are vanished or ignored due to orthogonality of $dt_\alpha$ (or $dt_\beta$) to $dt_0$ and of $dx_\gamma$ (or $dx_\sigma$) to $dx_n$. Due to the principle of CLT a differential equation of linear elements in according to  $(\ref{eq3})$, but with $\frac{\partial^2}{\partial t_3^2}$ instead of $\frac{\partial^2 }{\partial t_k^2}$, can be now added to  $(\ref{eq9})$ for fully describing the geodesic acceleration of deviation $\psi$ including rotation as well as linear translation:

\begin{equation}
\frac{\partial^2 \psi}{\partial {t_0}^2}- \psi\left ( \frac{\partial \varphi}{\partial t_0} \right )^2+\frac{\partial^2 \psi}{\partial {t_3}^2}=\frac{\partial^2 \psi}{\partial {x_n}^2}-\psi sin^2\theta\left ( \frac{\partial\varphi}{\partial x_n} \right )^2+\frac{\partial^2\psi }{\partial x_l^2};
\label{eq10}
\end{equation}

As in $(\ref{eq10})$  differentials $dt_3$ and $dt_0$, as well as corresponding covariant derivatives are locally orthogonal to each other, that we can group their second derivatives together as:

\begin{equation}
\frac{\partial^2 \psi}{\partial {{t_0}^+}^2}+\frac{\partial^2 \psi}{\partial {t_3}^2}=\frac{\partial^2 \psi}{\partial t^2};
\label{eq11}
\end{equation}

Where ${t_0}^+$ means an evolution toward the future. Similarly, due to a local orthogonality, for differentials $dx_l$ and $dx_n$, the second derivatives in 3D-space are also combined as:

\begin{equation}
\frac{\partial^2 \psi}{\partial {x_n}^2}+\frac{\partial^2 \psi}{\partial {x_l}^2}=\frac{\partial^2 \psi}{\partial {x_j}^2};
\label{eq12}
\end{equation}

In the result, two operations: i/ defining $\psi$ as a deviation parameter and ii/ the unification $(\ref{eq11})$ of two orthogonal time axes into the ordinary time $t$, almost hide the proper time $t_0$  and, simultaneously, reduce the 6D manifold into a 4D space-time. In principle, the vacuum circular polarization can separate time-like geodesic acceleration (as well as space-like one) in two opposite directions of evolution (or rotation). Before a separation of polarization there are symmetrical contributions of forward and backward evolutions in 3D-time and there is a symmetry between left-handed and right-handed contributions in 3D-space. However, due to transformation into 4D space-time and fixing the polarization, this symmetry is to be broken. For running the mechanism, qualitatively assuming that a Higgs-like tachyonic potential $V_H(\phi)=\lambda^2 \left [\phi^2-{\phi_0}^2\right ]^2$ induces a time-like centripetal force in 3D-time, where $\phi_0$ is the global vacuum field and $\lambda$ is the interaction constant. In ~\cite{Vo2} we estimated the major term of acceleration of the time-like deviation  $\psi$ in $(\ref{eq10})$:

\begin{equation}
\left ( \frac{\partial \varphi}{\partial t_0} \right )^2\psi=-\frac{V_H(\phi)}{\psi^2}\psi \Rightarrow \Lambda_T \psi=\left ( \frac{\partial \varphi}{\partial {t_0}^+} \right )^2\psi;
\label{eq13}
\end{equation}

Where $\Lambda_T$ is the effective time-like "cosmological constant" and the arrow means the breaking symmetry which is equivalent to selecting a single evolution toward the future accompanied by transformation from 6D time-space to 4D space-time. We propose that $(\ref{eq13})$ defines the major term of particle mass $m_0$.
\\The proper mass contains also space-like contribution corresponding to spinning in 3D-space which consists of P-even and P-odd terms. The squared P-even contribution is $m_s^2\sim \left ( k_n. \vec{s} \right )_{even}^2$, where $\vec{s}$ is the intrinsic spin of elementary particle and its normally oriented projection $s_l$ relative to the rotational plane $P_n$ can be both, left or right direction. In principle, electromagnetic or nuclear forces conserving the P-symmetry in 3D-space can be an inducing mechanism of the P-even contribution. Generally, without special polarization tool, the P-even contribution is to be hidden under the local geodesic acceleration condition in 3D-space:
\begin{equation}
\frac{\partial^2 \psi}{\partial {x_n}^2}-\psi sin^2\theta\left ( \frac{\partial \varphi}{\partial x_n} \right )^2\approx \frac{\partial^2 \psi}{\partial {x_n}^2}-\psi\left ( k_n.s_l \right )_{even}^2=0;
\label{eq14}
\end{equation}

The P-even mass contribution of the spinning term $m_s$ can be revealed in special polarization interactions. Instead of that, the P-odd contribution is a global proper polarization effect in term of the left-handed helicity, being observed universally in the weak interaction. We make a qualitative assumption that the breaking of symmetry in P-odd term may be caused by a vacuum potential of the global cosmological constant $\Lambda$ in 3D space:
\begin{equation}
\left ( \frac{\partial \varphi}{\partial x_n} \right )_{odd}^2\psi\Rightarrow \Lambda_L \psi=\left ( \frac{\partial \varphi}{\partial x_n^L} \right )^2\psi=-\hbar^{-2}\omega^2s_L^2\psi;
\label{eq15}
\end{equation}

Where $\Lambda_L$ is P-odd component of the space-like cosmological constant $\Lambda$ and the arrow again means the breaking symmetry by fixing a given helicity; $s_L$ is a normal spin projection to the plane $P_n$ equivalent to the left-handed helicity. The mass scale factor $\omega$ of P-odd contribution can be estimated from electroweak interference of leptons as $\omega \sim \alpha.G_F m_0^3;$ i.e. proportional to the fine structure constant $\alpha$ and Fermi constant $G_F$.
\\Finally, from $(\ref{eq10})$ we obtain the 4D space-time geodesic equation as follows:

\begin{equation}
-\frac{\partial^2 \psi}{\partial t^2}+\frac{\partial^2 \psi}{\partial {x_j}^2}=-\left [\Lambda_T -\left ( k_n.s_l\right )_{even}^2 -\Lambda_L \right ]\psi=-{\delta}_M^2 \psi;
\label{eq16}
\end{equation}

This nonhomogeneous squared differential equation is a transmission law of the function $\psi$ which characterizes a time-like curvature with a space-like adjustment. Therefore, Equation $(\ref{eq16})$ defines the emission of microscopic gravitational waves from gravitational sources ${\delta}_M^2\psi$. At variance with the traditional gravitational waves with a very small curvature, Equation $(\ref{eq16})$ contains a major strong time-like curvature which is inversely proportional to the microscopic wave function $\psi$. Moreover, because the gravitational sources are globally extended everywhere in 6D time-space, the interaction potential is to attach to the moving material point, described by the geodesic equation $(\ref{eq16})$. In the result, the speed of translational transmission of the wave phase is faster than the speed of light, which seems to ban observation. Nevertheless, the energy-matter following Equation $(\ref{eq16})$ is to transmit with an observable subluminal speed. Indeed, going on to rescale $(\ref{eq16})$ with the Planck constant and to make $\psi$ of a scale of the Compton length, we can find that the microscopic gravitational waves are identical to the well-known and well-observable quantum waves.

\section{Links with quantum mechanics}

Formulating quantum mechanical equations from a classical geodesic description, we adopt the quantum dynamic operators, such as: $\frac{\partial }{\partial t}\rightarrow i.\hbar\frac{\partial }{\partial t}=\widehat{E}$ and $\frac{\partial }{\partial x_j}\rightarrow -i.\hbar\frac{\partial }{\partial x_j}=\widehat{p_j}$. For the particle at rest, when $t\rightarrow t_0$ and $x_j\rightarrow x_n$, the operators are getting generators of proper masses: $i.\hbar\frac{\partial }{\partial t_0}=\widehat{E_0}=\widehat{m_0}$ and $-i.\hbar\frac{\partial }{\partial x_n}=\widehat{p_n}=\widehat{\delta m}$. This traditional procedure of quantum mechanics would be interpreted as a conversion between time and space which serves a mean for 4D space-time macroscopic observation of the microscopic gravitational waves transmitted in according to $(\ref{eq16})$.  Consequently, Equation $(\ref{eq16})$ leads to the basic quantum mechanical equation of motion:

\begin{equation}
-\hbar^2\frac{\partial^2 \psi}{\partial t^2}+\hbar^2\frac{\partial^2 \psi}{\partial x_j^2}-m^2\psi=0;
\label{eq17}
\end{equation}

Where the square mass term $m$ consists of the following components:   $m^2=m_0^2-\delta m^2=m_0^2-m_s^2-m_L^2$. In momentum representation Equation $(\ref{eq16})$ reads:

\begin{equation}
E^2\psi_p-\vec{p}^2\psi_p-m^2\psi_p=0;
\label{eq18}
\end{equation}

Equation $(\ref{eq18})$ describes subluminal motion of an elementary particle with energy $E$ and momentum $\vec{p}$. In comparison with the traditional expression of the rest mass, the present one includes an additional correction $\delta m$ associated with the contribution of the intrinsic spin in 3D-space. The P-even contribution $m_s$ linked with an external curvature of spinning in 3D-space can be compensated in according to the condition $(\ref{eq14})$ when only the linear translation along $x_l$ axis is taken in account for a laboratory frame observation. However, due to P-odd effect being observable in the weak interaction, the geodesic deviation of the material point by its spinning still induces a small non-zero mass scale factor $\left | \omega  \right |=m_L\ll m_s$ which proves a tiny internal curvature of our realistic 3D-space. In general, Equation $(\ref{eq17})$ is reminiscent of Proca equation of vector boson or the squared Dirac equation of lepton~\cite{Vo1}. In case of a scalar field or when there is no polarization analysis, $m \rightarrow m_0$, Equation $(\ref{eq17})$ turns to the traditional Klein-Gordon-Fock equation.
\\It is to mention that the local condition of geodesic deviation in 3D-space leads to:

\begin{equation}
\left ( \frac{\partial S}{\partial x_n} \right )^2=\left ( \hbar.sin\theta\frac{\partial \varphi}{\partial x_n} \right )^2=\frac{\hbar^2}{\psi}\frac{\partial^2 \psi}{\partial {x_n}^2}=-2mQ_B;
\label{eq19}
\end{equation}

Which is proportional to Bohm quantum potential $Q_B$ in ~\cite{Bo1}.
\\The existence of the spin term in $(\ref{eq17})$ is reminiscent of the Zitterbewegung of a free spinning electron (Schrodinger ZBW) ~\cite{Sc1}. In fact, when we describe in a laboratory frame a linear translation of a free particle by Equation $(\ref{eq17})$, the ZBW term is made almost hidden by the geodesic acceleration condition $(\ref{eq14})$ except a tiny P-odd term which is usually hard to observe. This implies a reason why ZBW is not observable experimentally without special measure of polarization or interference.
\\Suggesting that in the left side of $(\ref{eq9})$ the geodesic acceleration of deviation is restricted locally in 3D-time, one can derive the relation for time-energy indetermination:

\begin{equation}
\left | \Delta E \right |.\left | \Delta t \right |\geq \left | \Delta E_0 \right |.\left | \Delta t_0 \right |> \psi^{-1}\left | d\left ( i.\hbar\frac{\partial \psi}{\partial t_0} \right ) \right |.\left | dt_0 \right |=\left | i.\hbar \right |.d\varphi^2\geq \Delta \varphi_{min}^2\hbar\geq{0};
\label{eq20}
\end{equation}

Similarly, when the geodesic deviation acceleration is localized in 3D-space, the right side of $(\ref{eq9})$ leads to the space-momentum inequality:

\begin{equation}
\left | \Delta p \right |.\left | \Delta x \right |\geq \left | \Delta p_n \right |.\left | \Delta x_n \right |> \psi^{-1}\left | d\left ( i.\hbar\frac{\partial \psi}{\partial x_n} \right ) \right |.\left | dx_n \right |=\left | i.\hbar \right |.sin^2\theta d\varphi^2\geq \Delta \varphi_{min}^2\hbar\geq{0};
\label{eq21}
\end{equation}

Inequalities $(\ref{eq20})$ and $(\ref{eq21})$ can turn to equalities to zero only for the flat time-space of Euclidean geometry. For a non-zero curvature, it is proposed for $(\ref{eq21})$ an additional condition of  $sin^2\theta =1$, i.e. 3D-space quantization $\theta=(n+1/2)\pi$, equivalent to the cylindrical condition, and for both equations adopting $\Delta \varphi_{min}=\sigma\left(<\varphi>\right)=\sqrt{2\pi}$, where due to a statistical observability of the quantum indeterminism, $\sigma$ is a standard deviation of the mean value $<\varphi>=2\pi$ of an appropriate statistical distribution such as Poisson or Gaussian. In the result, we obtain the Heisenberg inequalities. This proves a direct link between internal time-space curvature of general relativity and Heisenberg inequalities in quantum mechanics.
Recalling that in ~\cite{Vo1} we derived the continuity equation of "a single particle" which in combination with the geodesic equation $(\ref{eq16})$ qualitatively allows to understand the physical reality of an individual particle in consistency with the quantum statistical interpretation and the context of wave-particle duality.

\section {Mass hierarchy of charged lepton generations}
In 4D space-time we assume that all charged leptons are to involve in the same basic time-like cylindrical geodesic evolution with an internal curvature of the time-like circle $S_1(\varphi^+)$, where $\varphi^+=\varphi({t_0}^+)$ is azimuth rotation in the plane $\{t_1,t_2\}$ about $t_3$ and the sign $+$ means an evolution toward the future. This universal feature determines the common properties of all charged lepton generations, except their mass hierarchy. Developing higher orders of curvature, we consider a generalized 3D time-like spherical system, described by nautical angles $\{\varphi^+,\theta_T,\gamma_T\}$, where $\theta_T$ is a zenith in the plane $\{t_1,t_3\}$ and $\gamma_T$ is another zenith in the orthogonal plane $\{t_2,t_3\}$. Coexisting in the same time-like cylindrical evolution $\varphi^+$, 4D observers see electron oscillating along a line-segment of the time-like amplitude $\Phi$, formulating one-dimensional comoving "volume": $V_1(\varphi^+)=\Phi=\psi.T$; where $T$ is the time-like Lagrange radius. For a homogeneity condition, equation of motion $(\ref{eq16})$ recalls the de Sitter solution.
In similar to the standard cosmological model, introducing a so-called microscopic cosmological model to the 3D time-like sphere, we consider $\Phi$ as the time-like microscopic Hubble radius and the wave function $\psi$ as the time-like scale factor. For instance, they are probably changeable during expansion of the Universe.
The highest order curvatures $C_n$ of $n-$hyper spherical surfaces are inversely proportional to $n-$ power of the time-like scale factor as $C_n\sim \psi^{-n}$. In particular, the energy density of electron correlates with its internal curvature as: $\rho_1=\epsilon_0/\psi$; where $\epsilon_0$ is assumed as a universal lepton energy factor. The mass of electron is determined as:
\begin{equation}
m_1=\rho_1V_1=\rho_1\Phi=\epsilon_0.T;
\label{eq22}
\end{equation}
The value $W_1=T$ is one-dimensional time-like Lagrange "volume" of electron. For muon and tauon except the common time-like cylindrical curved evolution $\varphi^+$, our 4D-observers can see some additional extradimensional curvatures come from simplest configurations of hyper-spherical surfaces $S_1(\theta_T)$ and $S_1(\gamma_T)$ or $S_2(\theta_T,\gamma_T)$. Those additional curvatures are external to 4D-observers as they are not involved in. The time-like "comoving volumes" $V_n(\Phi)$ with a fixed $\Phi$ are calculated as follows:
\begin{equation}
 V_n(\Phi)=\int_0^\Phi S_{n-1}(v)dv=\Phi.S_{n-1}(\Phi)=V_1S_{n-1};
\label{eq23}
\end{equation}
\\The energy distribution of n-hyper spherical configuration relates to electron density as: $\rho_n=\rho_1/\psi^{n-1}$. Therefore, the mass corresponding to $n-$dimensional configuration reads:
\begin{equation}
m_n=\rho_n.V_n(\Phi)=(\rho_1/\psi^{n-1})V_1S_{n-1}=W_1\rho_{n-1}S_{n-1};
\label{eq24}
\end{equation}
\\The simplest additional $S_1$ configuration is: $[S_1(\theta_T)+S_1(\gamma_T)]$, then the lepton mass of 2D time-like curved particle is:
\begin{equation}
m_2=W_1 \rho_1[S_1(\theta_T)+S_1(\gamma_T)]=\epsilon_04\pi.T^2 ;
\label{eq25}
\end{equation}
For the simplest additional $S_2(\theta_T,\gamma_T)$ configuration the corresponding lepton mass of 3D time-like curved particle is:
\begin{equation}
m_3=W_1 \rho_2 S_2(\theta_T,\gamma_T)=\epsilon_04\pi.T^3 ;
\label{eq26}
\end{equation}
 The equations of lepton mass are obtained here in the first approximation only, because the time-like curvatures of hyper-spherical surfaces would contain more precise terms. There are in  $(\ref{eq22}){,}(\ref{eq25})$ and $(\ref{eq26})$ two parameters: the lepton energy factor $\epsilon_0$ and the time-like Lagrange radius $T$ which would be determined, in principle, by experimental masses of two from three charged leptons and to use for predicting the mass of the third lepton. Instead of this, we assume a qualitative approach as an alternative scenario for estimation of Lagrange radius $T$ following the microscopic cosmological model:
 \\During the Big-Bang inflation the time-like scale factor $\psi$ increases exponentially with a time-like Hubble constant $ H_T=\sqrt {\Lambda_T}=7.764{*}10^{20}$ sec$^{-1}$, in according to $(\ref{eq16})$ and $(\ref{eq17})$. As the time-like Hubble radius $\Phi$ is fixed for an inflation instant of $\Delta {t_1}=1.926{*}10^{-20}$ sec. after $1.0$ sec. from the Big-Bang, their ratio $T$ decreases for $3.20{*}10^{-7}$ times from the initial unit value $T_0=\Phi/\psi_0=1$.  For the radiation dominant era of $\Delta {t_2}\sim 10^{11}$ sec. and for the next matter dominant era of almost all life-time of the Universe $\Delta {t_3} \sim 13.7$ Bill. years ($4.32{*}10^{17}$ sec.) it is assumed that: $\psi \sim t^{1/2}$ and $\sim t^{2/3}$, accordingly. As the Hubble radius increases faster, namely $\Phi \sim t$, the time-like Lagrange radius $T$  now steps up to the present value $T=\Phi/\psi=16.5$. For leptons born after the inflation era, we follow the anthropic principle to assume qualitatively that the Hubble radius of any quantum fluctuations should fit the contemporary value $\Phi$, while the scale factor $\psi$ being governed by a contemporary chaotic Higgs-like potential (in a such way, e.g. contracting) is also to meet the contemporary time-like Lagrange radius, then evolving to $T=16.5$ for today.
\\ In according to $(\ref{eq22})$ using the experimental electron mass for calibration and $T=16.5$, we found the lepton energy factor $\epsilon_0=31.0$ $keV$. Now equations $(\ref{eq25})$ and $(\ref{eq26})$ allow to predict the mass ratios and the absolute masses of muon and tauon (in $MeV$), as follows:
\begin{equation}
m_1:m_2:m_3=1:207.4:3421.5=0.511:106.0:1748.4;
\label{eq27}
\end{equation}
It is to compare with the experimental data of charged lepton masses from ~\cite{Ber1}:
\begin{equation}
m_e:m_\mu:m_\tau=1:206.76828:3477.2=0.510998928(11):105.6583715(35):1776.82(16);
\label{eq28}
\end{equation}
Here the electron mass is a calibration, while the calculated values of muon and tauon masses are found in $(\ref{eq27})$ in a good consistency with their experimental data in ~\cite{Ber1}, within $(0.32-1.60)\%$ of relative deviation, even for the first order of approximation. This result may serve a solution to the long-standing problem of charged lepton mass hierarchy, giving a possible answer on the puzzle of the "exact number three" of lepton generations. Naturally, the time-space symmetry leads to accepting a 3D-time geometry, the dimensions of which correlate strictly with the number of lepton generations.
\\If the expansion scenario of the microscopic cosmological model is applicable, the lepton masses in  $(\ref{eq22}){,}(\ref{eq25})$ and $(\ref{eq26})$ increase during cosmological expansion with different rates. Even being not able to see any change of electron mass, the 4D observers can measure the changeable masses of muon and tauon. It would open a unique window for experimental verification by measuring a time-variation of the mass ratios. In particular, $R_{21}=m_\mu:m_e=4\pi.T$ and the increasing rate is $T\sim t^{1/3}$. Let take a duration of data sampling roughly $\Delta t_s=10$ years, then the mass ratio increases accordingly by $\Delta R_{21}=\frac{4\pi T}{3}\frac{\Delta t_s}{13.7{*}10^9}=5.0{*}10^{-8}$. The estimation implies that a direct measurement of variation of the mass ratio needs to improve precision of both experimental data of muon and electron masses in $(\ref{eq28})$  by two orders more, before going on for comparison and observation of any change of their ratio.

\section{Conclusions}

Based on an original time-space symmetry we developed a geometrical dynamical model with a common intrinsic cylindrical curvature. We have shown that the extended 3D-time seems not to be a fictive subspace, but the time-like EDs are revealed in terms of the quantum wave function $\psi$ and the proper time $t_0$ (under the time-like rotational variable $\varphi$) which contribute to a full time-space evolution of microscopic particles. A duality was found that the second order differential equations in quantum mechanics, including Proca, squared Dirac and Klein-Gordon-Fock equations, are indeed the emission law of microscopic gravitational waves carrying a strong time-like curvature in a weakly curved 3D-space. The observed duality demonstrates a direct link between general relativity and quantum mechanics.
\\As the results, the local 3D geodesic acceleration conditions of deviation $\psi$ shed light on the origin of quantum phenomena such as: i/ Bohm quantum potential, ii/ Schrodinger ZBW of a spinning electron and iii/ Heisenberg inequalities. In particular, the derivation of quantum indeterminism is a strong evidence of internal curvatures of our realistic 4D space-time.
\\Number of lepton generations is assumed equal to the maximal time-like dimension, i.e. a 3D-time subspace:
Based on the common intrinsic cylindrical mode, we extend the time-like curvature to higher dimensional time-like hyper-spherical configurations to calculate the mass ratios of all leptons which are found in good quantitatively consistency with the mass hierarchy of charged leptons. It would serve a solution to the problem of "the exact number three" of lepton generations, simultaneously, proving the concept of time-space symmetry. The lepton masses can be changeable with different rates during a long period of cosmological expansion, therefore, a time dependence of their mass ratios would be a subject for experimental verification.
\\Observations in the present work imply a necessity of reformation of our basic concepts, such as the flatness of 4D space-time and the locality of the dynamical interactions in 4D stage. In this consideration, the quantum mechanics serves an effective 4D space-time holography for restoration of the physical world on a "lightcone" surface of the extended 6D symmetrical time-space.

\section*{Acknowledgment}
The author is grateful to A.V. Nguyen and A.K. Nguyen (Hanoi Institute of Physics) for their useful discussion. A hearty thanks is extended to B.N. Nguyen (Hanoi Thang-Long University) and M.L. Vo-Le${^,}$s for their technical assistance and precious encouragement.

\end{document}